\documentclass[%
aps,prl,twocolumn,superscriptaddress%,linenumbers
]{revtex4-1}
\usepackage{graphicx}% Include figure files
\usepackage{subfigure}
\usepackage{dcolumn}% Align table columns on decimal point
\usepackage{bm}% bold math
\usepackage{hyperref}% add hypertext capabilities
\usepackage{indentfirst}
\usepackage{braket}
\usepackage{amsmath}
\hypersetup{colorlinks=true, citecolor=blue, urlcolor=blue, linkcolor=blue}
\pdfoutput=1

\begin{document}
\title{Exotic Superfluid with Emergent Flux in a One-Dimensional Bose-Fermi Mixture}
\author{Qi Song}
\affiliation{Department of Physics and State Key Laboratory of Surface Physics, Fudan University, Shanghai 200433, China}
% \email{20110190009@fudan.edu.cn}
\author{Jie Lou}
\email{loujie@fudan.edu.cn}
\affiliation{Department of Physics and State Key Laboratory of Surface Physics, Fudan University, Shanghai 200433, China}
\author{Yan Chen}
\email{yanchen99@fudan.edu.cn}
\affiliation{Department of Physics and State Key Laboratory of Surface Physics, Fudan University, Shanghai 200433, China}
\affiliation{Shanghai Branch, Hefei National Laboratory, Shanghai 201315, P.R. China}

\date{\today}

\begin{abstract}
We find a novel chiral superfluid (CSF) phase in a one-dimensional Bose-Fermi Hubbard model with significant mass and density imbalance between the two species. 
In the CSF phase, bosons condensate at non-zero momentum $\pm 2\pi /L$ with chain length $L$. To capture the essential physics of this new phenomenon, we study an alternative simplified model that only features competition between single-fermion hopping and hopping of composite particles composed of a fermion and a boson. This model captures the low energy physics of the Hubbard model and hosts a robust CSF phase. Our unbiased numerical studies show that in the CSF phase, the local superfluid order parameter continuously rotates along the chain, indicating that time-reversal symmetry is spontaneously broken. This symmetry breaking generates an emergent flux in the background, effectively optimizing the system's ground-state energy. 
We provide a physical understanding at the mean-field level. Furthermore, we have explored the potential realization of this phase in cold-atom experiments.
\end{abstract}
\maketitle

% \emph{Introduction.---}
The quest for unconventional novel superfluids\cite{excitonSF2023,wang2021evidence,chiralSF2015,TSF2012,XpL2021,XPL2012,TSF2016,CSFpwave2016} has been a persistent research focus in quantum physics. In fermionic systems, significant theoretical and experimental efforts spanning decades have been dedicated to topological superfluids\cite{RevModPhys3HeA,topoSC2017} and the Fulde-Ferrell-Larkin-Ovchinnikov state with finite-momentum Cooper pairs\cite{FFLOreview2018}. In bosonic systems, superfluid phases exhibiting complex order parameters have been experimentally observed in two-dimensional optical lattices\cite{wang2021evidence,chiralSF2015,TSF2012}, and the Potts-nematic superfluidity, which exhibits spontaneous rotation symmetry breaking, has also been reported\cite{XpL2021}. Since Bose-Fermi mixtures (BFMs) incorporate both bosonic and fermionic degrees of freedom, the pursuit of novel superfluids within these mixtures presents a promising prospect.

The interaction between bosons and fermions is essential in condensed matter physics and can give rise to exotic states of matter. In conventional superconductors, phonon exchange between electrons triggers the formation of Cooper pairs\cite{schrieffer1983theory}. Over the last two decades, ultracold bosonic and fermionic atoms have emerged as remarkable platforms for simulating quantum many-body physics\cite{RevModPhys.80.885,Bloch2017QuantumSimulations, Tools}. Due to their high controllability, atomic mixtures have exhibited rich phenomena, such as double superfluids\cite{doi:10.1126/science.1255380, PhysRevLett.118.055301,doubleSF2016}, mediated interactions\cite{Nature568.61}, heteronuclear molecules\cite{PhysRevLett.93.143001,milczewski2023molecules,duda2023molecules}, and hybridized sound modes\cite{Soundexp2023PhysRevLett.131.083003, ZhigangWu2024PhysRevLett.132.033401}. Furthermore, topological $p$-wave superfluids in BFMs have garnered theoretical attention\cite{ZhigangWu2016, ZhigangWu2018}. Remarkably, spontaneous translational symmetry-breaking insulating states have recently been observed in two distinct solid-state BFM systems\cite{ruiz2023bose,zeng2023exciton,lagoin2023dual}. Consequently, BFMs are fertile ground for cultivating intriguing new superfluid phases. 

In this paper, we reveal a novel chiral superfluid (CSF) phase that spontaneously breaks time-reversal symmetry. In this phase, the bosonic superfluid order rotates along the one-dimensional (1D) periodic chain, creating an effective emergent flux. Consequently, bosons condense at non-zero momentum $\pm 2\pi /L$, where $L$ denotes the chain length. 
Initially, we find the CSF within the extensively studied Bose-Fermi Hubbard (BFH) model \cite{sowinski2019bfmReview,bfmPhaseDiagram2006,PhysRevLett.92.050401}, which describes a mixture of ultracold bosons and fermions confined within an optical lattice.  
Our investigation of the BFH model focuses on conditions characterized by a substantial imbalance in both the masses and densities of the two species, a regime that has received limited attention in previous studies\cite{sowinski2019bfmReview,bfmPhaseDiagram2006,PhysRevLett.92.050401,PhysRevA.77.023621,bfm2012MassImbalance,PhysRevA.77.023601}. 
Due to strong boson-fermion interactions, the low-energy physics of the system can be described by an effective model where composite-particle hopping competes with single-fermion hopping. Composite-particle hopping was introduced in the context of high-temperature superconductivity\cite{compositeHoppingPhysRevResearch.2.023291}. In our study, composite-particle hopping (a fermion hops together with a boson) captures the essence of boson-fermion correlations, which directly leads to the emergence of CSF. Based on this effective model, we provide a comprehensive understanding of the origin of CSF from two distinct perspectives, supported by concrete numerical evidence and mean-field analysis.

Remarkably, our system solely adopts the CSF as its ground state when an even number of fermions are present in the mixture. In contrast, it behaves as a normal superfluid state when the number of fermions is odd. This indicates that the chirality of this quantum system, protected by an energy gap, can be manipulated by adding or removing particles through the tuning of chemical potentials. This even-odd disparity underscores the topological nature of the CSF.
Finally, we discuss the potential for realizing and detecting the 1D CSF in cold-atom experiments.

%%%%%Figure %%%%
\begin{figure}
\centering
\includegraphics[scale=0.66,trim=0 0 0 0,clip]{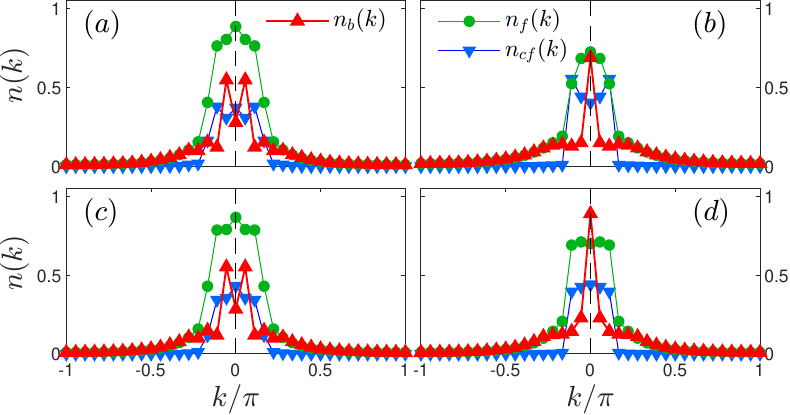}
\caption{\label{Fig0_nk_tcf_Ubf}Momentum distribution functions $n(k)$ obtained by DMRG for the Hubbard model (upper panels) and the effective model (lower panels) shows similar behavior. (a) $N_f=6=2N_b$ and (b)$N_f=5, N_b=3$ correspond to the BFH Hamiltonian(\ref{eqBFH}) at $t_{b}=0.25t_f,U_{bf}=-2.5t_f$. (c)$N_f=6=2N_b$ and (d)$N_f=5,N_b=3$ correspond to the composite-particle Hamiltonian(\ref{eqbf1}) at $t_{cf}=t_f$. For even $N_f$ (Left column), bosons condensate at $|k|=2\pi/L$ (positions of the two red peaks in $n_b(k)$). For odd $N_f$ (right column), bosons condensate at $k=0$. We set $U_{bb}=7.5t_f$ and $L=36$ for all cases.
}
\end{figure}
%%%%%%%%%%%%% 

\emph{CSF phase in the BFH Model.---} The standard BFH model, featuring an onsite boson-fermion interaction strength $U_{bf}$ within a 1D periodic chain, is expressed as follows:
\begin{eqnarray} \label{eqBFH}
H_{BFH}=&&-t_f\sum_{i} (c^{\dagger}_ic_{i+1}+H.c.)-t_b\sum_{i}(b^{\dagger}_ib_{i+1} +H.c.)\nonumber  \\
        &&+\frac{U_{bb}}{2}\sum_{i} n^b_{i}(n^b_{i}-1) + U_{bf}\sum_{i} n^b_{i}n^f_{i} 
\end{eqnarray}
where $c^{\dagger}_i$ ($b^{\dagger}_i$) creates a spinless fermion (boson) at site $i$, $t_f$ ($t_{b}$) is the tunneling amplitudes of fermions (bosons), and $n^b_{i}=b^{\dagger}_ib_i$ $(n^f_{i}=c^{\dagger}_ic_i)$ denotes the boson (fermion) number operator. The term
$U_{bb}$ suppresses double occupation of bosons and we set $U_{bb}=7.5t_f$ if not specified.
We employ both the density matrix renormalization group (DMRG)\cite{PhysRevLett.69.2863, ITensor} and Grassmann multi-scale entanglement renormalization ansatz (GMERA)\cite{lou2015combining} methods, two well-established unbiased numerical techniques based on tensor product ansatz to conduct calculations. 
Our primary focus is on the attractive interspecies coupling ($U_{bf}<0$), though similar results are obtained on the repulsive side \cite{SM}.

When the interspecies interaction is sufficiently strong, a composite fermion (CF) is formed through the pairing of a fermion and a boson, assuming equal densities of both species \cite{bfmPhaseDiagram2006,PhysRevLett.92.050401,PhysRevA.77.023621,bfm2012MassImbalance,Sachdev2005BFM}.
To introduce a density imbalance, we set the total number of fermions $N_f$ to be greater than the number of bosons $N_b$. This imbalance results in an excess of fermions without corresponding bosons to pair with, leading to strong scattering between fermions and composite particles and enhancing bosonic superfluidity. 

To avoid the scenario where bosons do not exhibit condensation, we ensure that the fermion density remains below 1/3 \cite{PhysRevA.77.023621}. Additionally, we incorporate significantly unequal hopping amplitudes, reflecting the inherent mass difference between heavy bosons and light fermions.
%%% U_{bf}
In the case of weak attraction, bosons occupy the normal superfluid phase with zero condensation momentum, while fermions are in the Luttinger liquid phase with a well-defined Fermi momentum. However, as the attractive strength increases, DMRG results indicate that bosons condense into the CSF phase at momentum $k=\pm 2\pi/L$ with even fermion number $N_f$ at $t_b=0.25t_f$, as illustrated in  Fig. \ref{Fig0_nk_tcf_Ubf}(a). 
In contrast, for odd  $N_f$ in Fig. \ref{Fig0_nk_tcf_Ubf}(b), bosons condensate at $k=0$, exhibiting conventional SF behavior. 
The distinct behavior between systems with even and odd parity recalls the behavior of fermions in a 1D periodic chain under the influence of magnetic flux \cite{Nakano_2000}. This observation inspired us to propose a straightforward effective model that captures the essential physics of this mixed system.

%%%%%Figure 1%%%%
\begin{figure}
\centering
\includegraphics[scale=0.75,trim=10 0 1 0,clip]{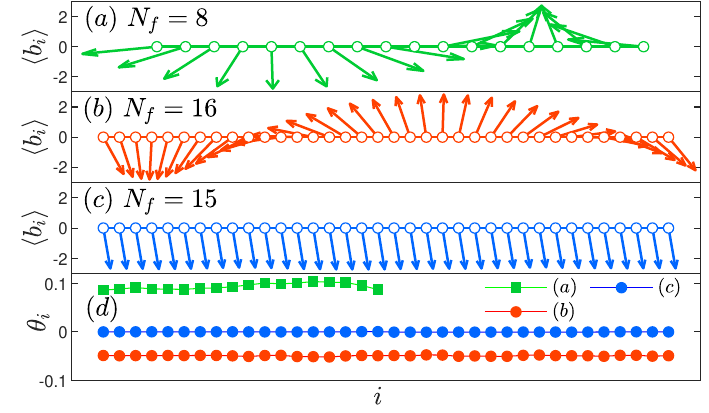}
\caption{\label{Fig1_b}(a)--(c):The boson superfluid order parameter $\langle b_i\rangle$ (arrows) measured on each site $i$ (open circles),
magnified ten times, with the horizontal/vertical component of the arrow corresponding to its real/imaginary part, respectively.
(d): The argument angle $\theta_i$ of the nearest-neighbor hopping term $\langle b_i^{\dagger}b_{i+1} \rangle$, corresponding to cases (a)--(c). 
Results are obtained by GMERA on the composite-particle Hamiltonian(\ref{eqbf1}) with parameters $t_{cf}=2t_f, U_{bb}=7.5t_f$, and the following particle numbers: (a)$N_f=8=2N_b$ for $L=18$; (b)$N_f=16=2N_b$ for $L=36$; (c)$N_f=15, N_b=8$ for $L=36$.
% Results are obtained at parameter setting $t_f=0.2,t_{cf}=0.4,U_{bb}=3.0$. 
% Here are the details regarding particle numbers: (a)$N_f=8=2N_b$; (b)$N_f=16=2N_b$; (c)$N_f=15, N_b=8$.
}
\end{figure}
%%%%%%%%%%%%%

%%%
Under the influence of boson-fermion attraction, a system characterized by an excess of fermions ($N_f>N_b$) can generally be viewed as a mixture of CFs and unpaired fermions. Whether bosons exhibit condensation behavior depends on the magnitude of attraction, as discussed in \cite{bfm2012MassImbalance}. 
%SACHDEV REFERENCE MOVE TO SOMEWHERE ELSE.
Considering the kinetic motion of bosons is strongly correlated with that of fermions, the resulting effective composite-particle Hamiltonian takes the form
\begin{eqnarray} \label{eqbf1}
H=&&-t_f\sum_{i} (c^{\dagger}_ic_{i+1} +H.c.) + \frac{U_{bb}}{2}\sum_{i} n^b_{i}(n^b_{i}-1)\nonumber \\
    && -t_{cf}\sum_{i} (c^{\dagger}_ib^{\dagger}_ic_{i+1} b_{i+1}+H.c.)
\end{eqnarray} 
where $t_{cf}$ is the amplitude of composite-particle hopping. 
Prior to collapse, bosons transition into a superfluid state for a nonzero $t_{cf}$ values.
As depicted in Fig. \ref{Fig0_nk_tcf_Ubf}(c-d), DMRG calculations for this composite-particle model reveal that the momentum distributions of bosons exhibit analogous even-odd disparities observed in the BFH Hamiltonian(\ref{eqBFH}). Notably, the Fermi surfaces of both fermions and CFs are discernible in both models. (Details of phase diagrams for the two Hamiltonians are given in the Supplementary Material\cite{SM}.)

\emph{Origin of CSF from the composite-particle model.---}
Distinct from DMRG simulations, which maintains a conserved particle count, GMERA operates within the grand-canonical ensemble framework, enabling direct measurement of the local superfluid order parameter $\langle b_i \rangle$.
In contrast to conventional superfluids, where condensation occurs at $k=0$, non-zero momentum condensation of the CSF leads to a complex local order parameter $\langle b_i \rangle$  that rotates continuously along the chain. The twisted angles $\theta_{i,i+1}$
(argument difference between $\langle b_i \rangle$ and $\langle b_{i+1} \rangle$ 
on two neighboring sites)
add up to $2\pi$ over a complete lattice period, as illustrated in Fig. \ref{Fig1_b}.
This chiral superfluid state spontaneously breaks time-reversal symmetry,
resulting in a unique and intriguing emergent flux state in the Bose-Fermi mixture.

We address the origin of CSF from two distinct perspectives, starting from the bare fermion limit ($t_{f} \gg t_{cf}$).
First, the kinetic term of composite particles ($t_{cf}$) suggests that a bosonic superfluid in the background induces projected "silhouette" composite particles. A CSF introduces an additional phase factor that fermions acquire during hopping. Effectively, this generates an "emergent flux" that further improves the system energy, analogous to the behavior of electrons hopping in a ring with a magnetic flux\cite{PhysRevB.40.9382}.
Based on this argument, we develop a mean-field consideration, resulting in an effective boson-free model that qualitatively aligns with our original Hamiltonian $H$ in the bare fermion limit.
Second, the fermionic response to the bosonic emergent flux is crucial.
As the CSF emerges, it profoundly modifies the fermionic momentum distribution. Numerical calculations reveal that the degeneracy of the fermionic Fermi surface (two Fermi points) is lifted simultaneously when the time-reversal symmetry of the superfluid is broken.
This lifting of degeneracy is accompanied by a redistribution of the momentum distribution of composite particles, which becomes more concentrated around small  $k$ values, further optimizing the system's energy. The collaborative effect between the fermionic and bosonic sectors stabilizes the CSF state.

\emph{CSF and Fermions parity.---}
Furthermore, our system exhibits an intriguing sensitivity to the parity of fermions. Specifically, when the number of fermions $N_f$ is even, the ground state is in the CSF phase where bosons condensation occurs at $k_0=\pm 2\pi /L$, either clockwise or counterclockwise, depending on the sign of $k_0$. However, when $N_f$ is odd, bosons condense at $k=0$, similar to a normal superfluid. 
In both cases, the superfluid phase persists to an intense competition region $t_f \approx t_{cf}$. 
Numerical results from the long-range boson correlation $G_b(x) = \langle b_i^{\dagger}b_{i+x} \rangle$
and momentum distribution $n_{b}(k)$ show that superfluidity emerges as soon as $t_{cf}$ is turned on, such as ($t_{cf}=10^{-3}t_f$). 
In the weakly coupled case, with a boson superfluid as the background, we understand that the free hopping fermions can gain additional energy through the $t_{cf}$ 
term in Eqn. \ref{eqbf1}, by projecting out silhouette composite particles.
When $N_f$ is even, the energy contribution from the $t_{cf}$ term can be further optimized by allowing superfluid order parameter $\langle b \rangle$ to rotate and acquire chirality.
In such cases, the hopping amplitude $\langle b_i^{\dagger}b_{i+1} \rangle$ becomes a complex number due to the phase difference of $\langle b \rangle$  between neighboring sites. Consequently, the expression 
$c_i^{\dagger}c_{i+1} b_i^{\dagger}b_{i+1}$ should be regarded as
$ c_i^{\dagger}c_{i+1}  \langle b_i^{\dagger}b_{i+1}  \rangle$.
Numerical results confirm that the measurement of $\langle b_i^{\dagger}b_{i+1}\rangle$ on different bond 
($i,i+1$) exhibit nearly identical arguments and moduli,
as shown in Fig. \ref{Fig1_b}(d). 
This observation suggests that a CSF generates an additional phase factor $e^{i\theta_{i}}$ in the composite-particle hopping term.

%%%%%%%%%% Fig %%%%%%%
\begin{figure}
\centering
\includegraphics[scale=0.9,trim=5 0 0 0,clip]{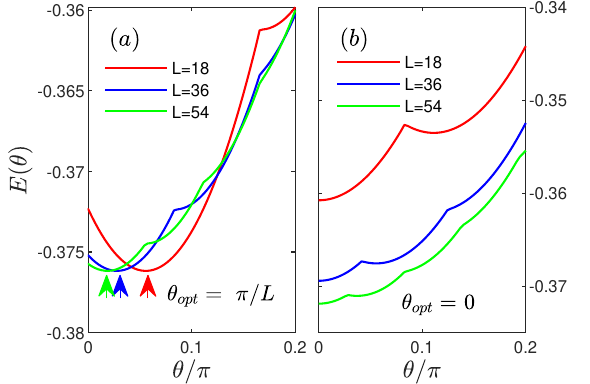}
\caption{\label{Fig4_MF}The ground state energy $E(\theta)$ of the mean-field Hamiltonian $H_{M F}$ as a function of the argument $\theta$ in $L=18,36,54$ systems when $t^{\prime}_{cf}=2t_f$. (a) Even fermion parity. The fermion filling factor is fixed at $n_f =N_f/L= 4/9$. There is a prominent peak on $\theta=0$ where no magnetic flux is threaded through the system. And $E(\theta)$ reaches its absolute minimum at $\theta_{opt}=\pi/L$. (b) Odd fermion parity. $N_f$ are $7,15,23$ for $L=18,36,54$ respectively. No peak shows on $\theta=0$ contrary to the even case in the main panel. And $E(\theta)$ reaches its absolute minimum at $\theta_{opt}=0$  }
\end{figure}

\emph{Explanation from effective flux.---}
The above finding enlightens us to trace back to the "flux phase" problem
\cite{WIEGMANN1988103,10.5169/seals-116403,1992Fluxes,1996On} which provides insight into
why magnetic flux can lower the ground-state energy of fermions. 
Extensive studies have shown that the energy of electrons can be improved by the presence of magnetic flux penetrating the lattice,
both in one-dimensional (1D) systems \cite{Nakano_2000} and in higher dimensions \cite{PhysRevLett.63.907,PhysRevB.41.9174,PhysRevLett.73.2158}.
In the specific case of spinless fermions in 1D,
rigorous theoretical works \cite{PhysRevLett.111.100402,PhysRevB.97.125153,Zheng_2020} have established that the optimal total flux phase $\Phi=\pi$ when $N_f$ is even, $0$ otherwise.

%%%%%Figure 3%%%%
\begin{figure}
\centering
\includegraphics[scale=0.8,trim=0 0 0 0,clip]{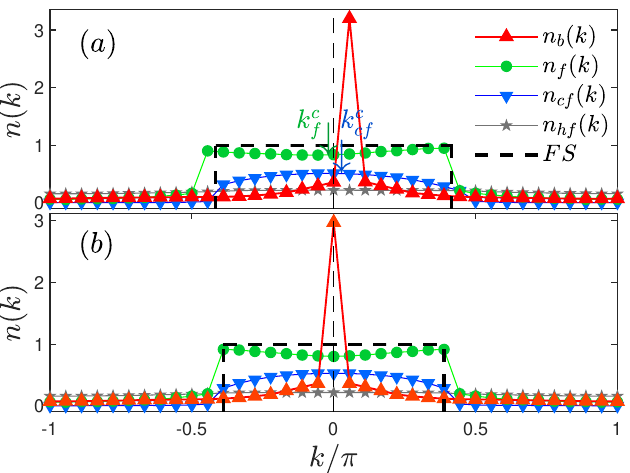}
\caption{\label{Fig3_GMERA_nk}Momentum distribution functions $n(k)$ for $L=36$ system with boson filling factor $n_b \approx 1/4$, 
obtained by GMERA on the composite-particle Hamiltonian(\ref{eqbf1}). "FS" refers to the symmetric Fermi surface of original fermions. 
(a) For $N_f=16$ (even), bosons condensate at $|k_0|=2\pi/L$ (position of the blue peak in $n_b(k)$). 
$k^c_f=\sum_k n_f(k)\times k /N_f=-0.0122\pi$ is the center of original Fermi sea, 
whereas $k^c_{cf}=\sum_k n_{cf}(k)\times k /N_{cf}=0.0075\pi$ is the center of Fermi sea for composite fermion, where $N_{cf}=\sum_k n_{cf}(k)$. 
(b) For $N_f=15$ (odd), bosons condensate at $k=0$. The parameters are chosen as $t_{cf}=2t_f,U_{bb}=7.5t_f.$ 
Fourier transforms of $G_b(i,j) = \langle b_i^{\dagger}b_{j} \rangle$, $G_f(i,j) = \langle c_i^{\dagger}c_{j} \rangle$, $G_{cf}(i,j) = \langle c_i^{\dagger} b_i^{\dagger} c_{j}b_{j} \rangle$ and $G_{hf}(i,j) = \langle c_i^{\dagger} b_i c_{j}b_{j}^{\dagger} \rangle$ give respective momentum distributions
$n_b(k),n_f(k),n_{cf}(k)$ and $n_{hf}(k)$. The black dashed lines guide the eyes at $k=0$.
}
\end{figure}
%%%%%%%%%%%%%
% {\bf settings}

Next we apply a mean-field approximation to our Hamiltonian (\ref{eqbf1}) to acquire a corresponding boson-free model
\begin{flalign}\label{eqMF}
\nonumber H_{M F}=-t_f\sum_{i}(c^{\dagger}_ic_{i+1} +H.c.)-t^{\prime}_{cf}\sum_{i} (c^{\dagger}_ic_{i+1} e^{i\theta}+H.c.)
\end{flalign}
The second term can be intuitively understood through our mean-field ansatz, treating the influence of bosons classically as a pure phase factor $e^{i\theta}$. Therefore, we substitute the hopping term  $c_i^{\dagger}b_i^{\dagger} c_{i+1} b_{i+1}$ with $c_i^{\dagger}c_{i+1} e^{-i\theta}$. This model highlights the two competing hopping processes:
the former stands for original free hopping, whereas the latter depicts hoppings experiencing a lattice flux.
This simple model can be solved straightforwardly. As shown in Fig.~\ref{Fig4_MF}, 
the ground state energy $E$ exhibits a periodically varying function superimposed on a parabolic background, illustrating the competition between two distinct hopping processes.
Remarkably, a significant difference is also observed between systems with even (left panel) and odd (right panel) fermion parity:
when $N_f$ is even, the optimal $\theta$ equals to $\pi/L$. In contrast, when $N_f$ is odd, the optimal $\theta$ takes the value of $0$.
This observation from the mean-field model is similar to the results obtained in the BFM.
However, it is crucial to note that  $\theta_{i}$, the argument of the complex $\langle b_i^{\dagger}b_{i+1}\rangle$ shown in Fig. \ref{Fig1_b}(d), averages out to $0.0940\pi$ for $L=18$ or $-0.0491\pi$ for $L=36$,
deviating from the optimal angle of $\pm \pi/L$ predicted by mean-field theory.
This deviation is attributed to the homogeneous condition of our 1D periodic lattice.

\emph{Shift of Fermi surface.---}
Interplay and correlations between different species of particles lead to the shift of the Fermi surface. We focus on the correlation function and momentum distribution of involved particles, influenced by the effective flux. When $N_f$ is odd, the conventional bosonic superfluid serves as a flux-free background, the correlation function of projected composite fermions $G_{cf}(i,j) = \langle c_i^{\dagger} b_i^{\dagger} c_{j}b_{j} \rangle$ shows similar 
long-range behavior as that of an original fermion.
As demonstrated in Fig. \ref{Fig3_GMERA_nk}, two clear Fermi surfaces (points) of composite particles have the same momentum as those of spinless fermions.
In contrast, when  $N_f$  is even, the ground state exhibits a two-fold 
degeneracy: bosons condensate at either $+2\pi /L$ or $-2\pi/L$,
time-reversal symmetry of the system is spontaneously broken.
Intriguingly, this degeneracy lift simultaneously affects the Fermi surface of a bare fermion.
The selection of the last filled momentum is determined by the direction of the effective flux, resulting in an asymmetric bulk of the Fermi surface.
Consequently, the Fermi surface of the projected composite fermion shifts towards $k=0$, optimizing the system's overall energy.
In this uncorrelated limit, this projection can be rationalized as follows:
$n_{cf}(k)=1/L \sum_{i,j} G_{cf}(i,j)e^{ik(x_i-x_j)} 
\approx 1/L \sum_{i,j} G_f(i,j)G_b(i,j)e^{ik(x_i-x_j)}
\approx 1/L \sum_{i,j} G_f(i,j)|A|e^{i(k+k_0)(x_i-x_j)}.$
% \begin{eqnarray} \label{eqbf2}
% n_{cf}(k)&&=1/L \sum_{i,j} G_{cf}(i,j)e^{ik(x_i-x_j)} \nonumber \\
% &&\approx 1/L \sum_{i,j} G_f(i,j)G_b(i,j)e^{ik(x_i-x_j)}\nonumber \\ 
% &&\approx 1/L \sum_{i,j} G_f(i,j)|A|e^{i(k+k_0)(x_i-x_j)}
% \end{eqnarray}
Here, $G_b(i,j)=|A|e^{ik_0(x_i-x_j)}$ represents the boson correlation for superfluid condensates at momentum $k_0$.

By carefully comparing DMRG and GMERA results, we confirm that the CSF phase is energetically favorable. DMRG calculations often maintain degeneracy and symmetry, as evident in Fig.~\ref{Fig0_nk_tcf_Ubf}. 
Conversely, GMERA outcomes are prone to symmetry-breaking.
In Fig.~\ref{Fig3_GMERA_nk}, we present the momentum distribution derived from GMERA calculations. Please refer to the Supplementary Material \cite{SM} for further details.

\begin{figure}
\centering
\includegraphics[scale=0.72,trim=2.5 0 0 0,clip]{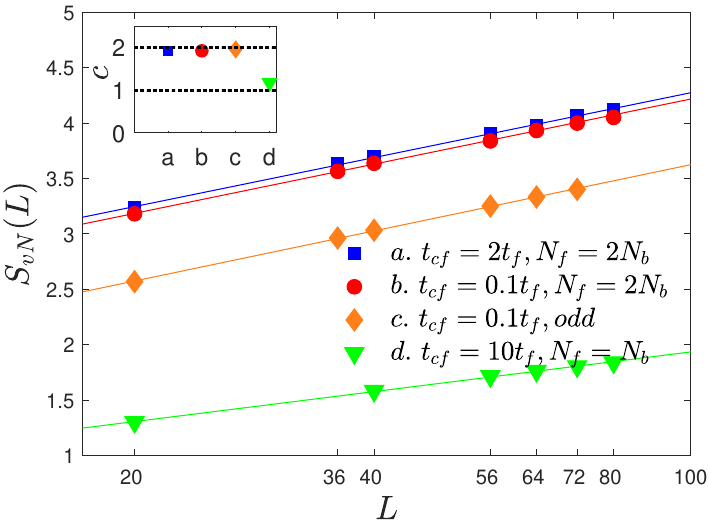}
\caption{\label{Fig5_cc_Nf_2Nb}The entanglement entropy $S_{vN}$ as a function of $L$ at different values of $t_{cf}/t_f$ and fermion parity in the semilogarithemic plot, obtained by DMRG. Boson and fermion numbers are $N_f=2N_b=L/2$ in 'a' (blue solid squares) and 'b' (red solid circles), $N_f=L/2-1, N_b=L/2$ in 'c' (orange solid diamonds) and $N_f=N_b=L/2$ in 'd' (green solid triangles). Inset shows the results of central charge.}
\end{figure}
%%%%%%%%%%%%%  

\emph{Comments about strong correlations.---}
Although the previous explanation is based on the uncorrelated limit, the CSF phase remains remarkably stable up to $t_f \approx t_{cf}$. 
A noteworthy difference emerges between $\langle c_i^{\dagger} b_i^{\dagger} c_{i+1}b_{i+1} \rangle$ and
$\langle c_i^{\dagger} b_i c_{i+1}b_{i+1}^{\dagger} \rangle$ that grows as $t_{cf}$ increases.  This distinction suggests a deviation from the uncorrelated scenario since a straightforward uncorrelated superfluid background fails to distinguish between these two terms.
Additional evidence can be found in the momentum distribution $n(k)$ of the intermediate region. The singularity of the Fermi surface for spinless fermions and composite particles is significantly altered by strong correlations, manifesting as convex and concave features in $n(k)$  near $k=0$ due to strong scattering.

Intriguingly, the central charge obtained from entanglement entropy in DMRG calculations remains consistently $\approx 2$, as illustrated in Fig. \ref{Fig5_cc_Nf_2Nb}. This observation indicates that, despite the projected particle's contribution to optimizing the system's total energy, it resembles the original particles. Global excitations are contributed by bosons superfluid and fermions Luttinger liquid.
It is worth mentioning that the condensation of a bosonic superfluid at a nonzero momentum has been previously explored in theoretical works, including those studying systems with correlated hopping \cite{PhysRevLett.108.225304, Keilmann2010StatisticallyIP}, as well as in the experimental realization of a two-dimensional topological chiral superfluid in optical lattices \cite{wang2021evidence,chiralSF2015, TSF2012}.
Nevertheless, the microscopic origin of the one-dimensional CSF in our system is fundamentally different. It arises directly from the interplay between two species of fermions and serves as a background field to optimize the system's overall energy.

 \emph{Discussions.---}
With the rapid advancement of quantum simulation techniques \cite{RevModPhys.80.885, Bloch2017QuantumSimulations, Tools}, a diverse array of interacting BFMs has been experimentally explored \cite{PhysRevLett.93.143001, milczewski2023molecules, duda2023molecules, 6Li7LiScience2001, PhysRevA.79.021601, PhysRevLett.93.183201}. Our system, characterized by a density imbalance between the two species, can be readily realized by loading unequal numbers of ultracold bosonic and fermionic atoms into a 1D optical lattice. Tunable Feshbach resonances provide precise control over the short-range boson-fermion interaction strength \cite{FeshbachRevModPhys2010, PhysRevLett.93.143001, PhysRevLett.93.183201}. Employing light fermions and heavy bosons, such as $^6$Li-$^{23}$Na\cite{PhysRevLett.93.143001} and $^6$Li-$^{41}$K\cite{6Li41K2018Feshbach,6Li41K2020JianweiPan}, naturally results in an inequality between the tunneling amplitudes of the two species. Furthermore, the tunnelling amplitude can be finely tuned by adjusting the depth of the species-dependent lattice potential\cite{SpinDependentLattice2004,SpinDependentLattice2011nature}.
%%% species-dependent lattice potential
Experimentally, periodic boundary conditions have been achieved in a 1D ring-shaped optical lattice with finite sites\cite{Franke_Arnold_2007,PhysRevLett.95.063201,Houston_2008,amico_2014_superfluid,LuigiAmico2022_ringLattice,Wayne2023ring}.
The characteristic signature of the CSF phase can be observed through widely utilized time-of-flight (TOF) experiments, which restore phase coherence information throughout the entire lattice\cite{2002TOF, RevModPhys.80.885}. 
The density distribution captured in TOF images corresponds to the momentum distribution of the trapped atoms.
During TOF expansion, the fermionic atoms influenced by an effective flux induced by the bosonic CSF will exhibit a doughnut-shaped pattern (analogous to the right panel of Fig.5 in Ref.\cite{LuigiAmico2022_ringLattice} and Fig.4 in Ref.\cite{Wayne2023ring}). 
This pattern contrasts with the bell-shaped pattern observed when bosons are in a normal superfluid state and fermions experience no effective flux \cite{LuigiAmico2022_ringLattice}.

% \emph{Conclusions.---}
 Our study reveals a novel many-body effect in a 1D BFM system, where the CSF phase emerges as the ground state and spontaneously breaks TSR to optimize the energy of the whole system.
Concrete numerical evidence and theoretical analysis provide a comprehensive understanding of the origin of CSF. Notably, the ground state can be tuned between the CSF and a conventional superfluid by  adjusting chemical potentials. Furthermore, we demonstrate the experimental feasibility of realizing this innovative state. Our discovery opens up opportunities to explore new quantum phases of matter.

% \section{Acknowledgements}\label{Acknowledgements}
\begin{acknowledgments}
\emph{Acknowledgements.---}
We thank Ting-Kuo Lee and Kai Chang for their fruitful suggestions. This work is supported by the National Key Research and Development Program of China Grant No. 2022YFA1404204, and the National Natural Science Foundation of China Grant No. 12274086. 
\end{acknowledgments}

\bibliography{bfm}
\end{document}